# Optical Scatter Imaging using Digital Fourier Microscopy

*K. Y. T. Seet[1,2], P. Blazkiewicz[1], P. Meredith[1], A. V. Zvyagin[1,2]*

[1]Centre for Biophotonics and Laser Science, School of Physical Sciences, The University of Queensland, St Lucia, 4072, Australia

[2]School of Information Technology and Electrical Engineering, The University of Queensland, St Lucia, 4072, Australia

Email: seet@physics.uq.edu.au,  Fax: +61 7 3365 1242, Tel: +61 7 3346 9727

**Abstract**

An approach reported recently by Alexandrov et al. [1] on optical scatter imaging, termed digital Fourier microscopy (DFM), represents an adaptation of digital Fourier holography to selective imaging of biological matter. Holographic mode of recording of the sample optical scatter enables reconstruction of the sample image. Form-factor of the sample constituents provides a basis for discrimination of these constituents implemented via flexible digital Fourier filtering at the post processing stage. Like in the dark-field microscopy, the DFM image contrast appears to improve due to the suppressed optical scatter from extended sample structures. In this paper, we present theoretical and experimental study of DFM using biological phantom that contains polymorphic scatterers.

## 1. Introduction

The intricate morphology of biological matter presents a challenge to biomedical imaging. Typically, it is manifested by low visibility features of interest on a crowded background of biological or cellular tissue. Therefore, the suppression of obscuring backgrounds in favour of features (termed gating), e.g. a cell organelle on the background of a live cell culture, represents an important research focus in biomedical imaging. There are several strategies to address this problem. In confocal microscopy, out-of-plane morphological layout of a biological specimen is suppressed in favour of the image plane of interest situated at the focal plane [2]. Another approach widely used in biomedical imaging is the labelling of biological constituent under study, this enables efficient discrimination relative to a non-labelled background [3].

Recently, a novel gating approach, termed optical scatter imaging (OSI), was successfully demonstrated by Boustany *et al.* [4]. This approach can be classified as Fourier filtering, and is applicable for thin low-scattering biological samples. It was demonstrated that OSI was sensitive to the minute size variations of the specimen constituents. In the transfer plane of the OSI imaging train terminated by a CCD camera, a field stop and variable diaphragm were placed to record two consecutive images. The first image was recorded with the diaphragm wide-open, this permitted collection of large-angle optical scatter. The second image was recorded with a reduced diaphragm opening that permitted collection of the small-angle optical scatter. The ratio of the two images allowed clear observation of cell apoptosis, a process which results in the swelling of sub-micron cell organelles. These same changes were undetectable in phase-contrast microscopy [4]. Also, the preferential orientation of the major constituents of the biological specimen has been demonstrated to be uniquely detectable using modified OSI [5].

The power of OSI is derived by the following properties: Firstly, OSI belongs to a family of dark-field microscopy, since it contains a field stop at the centre of the transfer plane [6]. Here a major portion of the optical radiation scattered in small angles (small-angle optical scatter), predominantly originating from the extended structures, is blocked by the field stop. As a result, the optical scatter is redistributed to boost its large-angle wing, with the effect of increased contrast of the small objects. The removal of the small-angle excess radiation from the detection channel affords multi-fold increase of the illumination power on the sample before the CCD camera saturates. Secondly, the sample optical scatter is controlled in OSI via operation of the diaphragm opening in the transfer plane and can be thought as implementation of Fourier filtering in hardware. The application of this filtering has enabled the discrimination of subwavelength-size scatterers [4]. This property is equivalent to the Fourier filtering that further enhances the OSI imaging performance and also provides additional information on the specimen scattering constituents. Despite these merits, OSI has several limitations. Among those, the optical scatter is controlled in hardware, which is slow in its current implementation. Also, the choice of the field stop and diaphragm shapes are hitherto limited to circular geometries, whereas angular scatter can be profoundly asymmetric [5]. Most importantly, phase of the optical scatter is lost in the OSI detection, this ultimately precludes optimal image processing.

In order to overcome the limitations of OSI, a new approach has been introduced to record and control the optical scatter based on digital Fourier holography, which we call digital Fourier microscopy (DFM) [1]. In DFM, the full optical scatter information is preserved since optical field amplitude and phase are recorded on the transfer plane using a CCD camera. This recording method enables efficient and adaptable Fourier filtering of the acquired holograms followed by sample image reconstruction in software [7, 8]. The most intense small-angle optical scatter is considerably de-emphasised in favour of the large-angle scatter by virtue of the limited dynamic range of a CCD camera. This underpins the field-stop action of DFM and can be, to some extent, controlled by the reference and object optical illumination. OSI merits such as small object contrast enhancement and Fourier filtering capacity, are preserved in DFM. At the same time, the main OSI limitations derived from its hardware implementation of the Fourier filtering are eliminated by transferring the optical image processing to the post-processing stage.

The use of digital Fourier holography [1] to perform optical scatter imaging represents a major departure from the previously reported approaches. A comprehensive account on principles, methods and algorithms of digital holography has been given by Yaroslavsky [9]. The maturity of digital sensor and computing technology has provided a significant impetus to new developments in the field of digital holography [10], this has a distinctive emphasis on biomedical imaging applications. It has been demonstrated that the state-of-the-art digital holography techniques are capable of 3D-imaging of live biological specimen, e.g. cells, at sub-video frame rate with subwavelength axial resolution [11, 12]. The use of holographic filters for imaging of biological tissue has been also recently demonstrated by Yu *et al.* [13]. Until recently, the use of Fourier holography has been limited to pattern recognition and optical computing applications [6, 9]. Most recently, Sebesta and Gustaffson have demonstrated the promise of digital Fourier holography for delineation of the embedded dielectric objects based on their refractive indices [14]. In this work, we present a theoretical and experimental study of the key properties of digital Fourier microscopy, including its Fourier filtering capacity and small object contrast enhancement.

The paper is organized as follows: In Section 2, we will introduce digital Fourier microscopy and detail its basic theoretical model, including comparative analysis of image contrast of DFM, and direct imaging methods. Details of the DFM experiments are presented in Section 3. Experimental results and discussion are addressed in Section 4 and 5, respectively, after which we will draw several concluding remarks.

## 2. Theory

### 2.1 Digital Fourier Microscopy

A simplified diagram of digital Fourier microscopy is shown in figure 1. An object is placed at the front focal plane of a Fourier lens $L_F$, of focal length $f_F$, that contains a CCD camera at its back focal plane, commonly called the transfer plane. A collimated monochromatic light beam is equally split between reference and object arms. In the reference arm the light beam is coupled into a single-mode optical fibre OF whose distal end is positioned at the front focal plane. Light emerging from this fibre is collimated by $L_F$, and propagates as a quasi plane wave at an angle to the optical axis. In the object arm, the incident light field is disturbed by an object, which is describable in the framework of light scattering. This optical disturbance is projected onto the transfer plane by means of $L_F$. Assume that the object is optically thin, i.e. its thickness is less than the so-called Rayleigh range determined as $2\lambda f_F^2 / (\pi d_0^2)$, where $\lambda$ is the light wavelength in the object medium, $d_0$ is the pupil diameter of the optical system. It is then reasonable to disregard the axial extension of the object for a moment and describe it as a two-dimensional (2D) function $f(x', y')$, termed object function, in the $x' - y'$ plane, with an origin chosen at the reference point-source, i.e. the reference fibre end (see figure 1). Although $f(x', y')$ can be generally a complex function representing a thin attenuating phase plate, we assume $f(x', y')$ real, i.e. purely attenuating. The object is assumed to be centred at $(x'_0, y'_0)$. The object and reference waves interfere at the transfer plane. The un-scattered portion of the illumination is omitted from the current theoretical analysis. The resultant interference intensity distribution on the transfer plane ($x - y$) forms a hologram, this represents a constant intensity background, devoid of information content, modulated by a 2D-holographic function $I(x, y)$, with an origin at the intersection of the transfer plane and optical axis. In the considered case of Fourier holography, this intensity distribution takes a particularly simple form [15]:

$$I_F(x, y) \propto F\left(\frac{x}{\lambda f_F}, \frac{y}{\lambda f_F}\right) \cos\left[\frac{2\pi}{\lambda f_F}(xx'_0 + yy'_0)\right], \tag{1}$$

where $F$ symbolizes a 2D-spatial Fourier transform of $f(x', y')$ defined as

$$F\left(\frac{x}{\lambda f_F}, \frac{y}{\lambda f_F}\right) = \iint f(x', y') \exp\left[-2\pi j\left(x'\frac{x}{\lambda f_F} + y'\frac{y}{\lambda f_F}\right)\right] dx' dy'. \tag{2}$$

The inverse Fourier transform is defined as:

$$f(x', y') = \frac{1}{\lambda^2 f_F^2} \iint F\left(\frac{x}{\lambda f_F}, \frac{y}{\lambda f_F}\right) \exp\left[2\pi j\left(x'\frac{x}{\lambda f_F} + y'\frac{y}{\lambda f_F}\right)\right] dx dy. \tag{3}$$

In both cases, the integration is assumed over the entire 2D space. As is evident from the observation of Eq. (1), the optical field disturbance $f(x', y')$ at the object plane can be reconstructed by performing a 2D-Fourier transform that is conveniently implemented using digital recording and processing technologies. Examination of Eq. (1) shows that the hologram is modulated at a constant spatial frequency $\nu_0 = \sqrt{x_0'^2 + y_0'^2}/(\lambda f)$ provided the thin object is situated at the object plane. In other words, the Fourier hologram represents a family of equally spaced ($1/\nu_0$) linear fringes. Conversely, the hologram of an off-plane thin object is modulated by a family of curved interference fringes whose separation increases with the distance from the origin. It represents the basis for a 3D-reconstruction of a set of thin objects, or thick sample surface profile [16]. Demonstration of the 3D-reconstruction is beyond the scope of this paper, although it represents an interesting topic of future research on DFM. In the following sub-sections, we will make several observations regarding imaging properties of DFM.

*2.2 Fourier filtering*

Eq. (1) shows that the 2D-Fourier transform $F\left(\dfrac{x}{\lambda f_F}, \dfrac{y}{\lambda f_F}\right)$, describes an envelope of the hologram determined by the object form-factor (shape, size), and relative refractive index, i.e. a ratio of refractive index of the object to that of the background medium. In the case of an ensemble of objects, i.e. a composite object, in the vicinity of the object plane, $F$ is a sum of each constituent's 2D-Fourier transform, each of which is centred about the origin at the transfer plane. Application of a 2D-mask determined for the constituent of interest $F_0$, to the recorded hologram, e.g. implemented as a product of the mask and hologram, followed by hologram reconstruction, will predominantly pick-up those constituents whose Fourier transform is conformal with $F_0$. This property central to the DFM merits will be further addressed in the experimental context of this paper.

*2.3 Contrast*

**a. Analytical model**

DFM can offer contrast enhancement, as compared to conventional bright-field microscopy. Consider the bright-field microscopy and DFM principal configurations as shown in figures 2(a) and 2(b), respectively. In the bright-field microscopy, an object situated at the object plane is illuminated with a plane wave of power $P$ and the image is recorded by a CCD camera placed at the image plane. Specifically, the imaging system is assumed to have a magnification of unity. If the illumination beam could then be eliminated from the detection process, as in case of the dark-field microscopy, the intensity distribution at the CCD sensor would directly relate to the object function $f$, discretely sampled by the CCD camera, $f(m,n)$. $f(m',n')$ is a discrete representation of the object function (prime identifies indexing associated with the object plane), whose indexing is identical to $f(m,n)$: $1 \le m', m \le M, 1 \le n', n \le N$, where $M$ and $N$ stand for the total number of pixels of the discrete function along the $x$ and $y$- axes, respectively. Otherwise, the illumination beam fills the entire area of the CCD camera. The bright-field signal, subscripted as BF, recorded by the CCD camera can be expressed in terms of number of photoelectrons at each pixel indexed by $m$ and $n$:

$$\phi_{BF}(m,n) \cong \rho\tau\frac{P}{MN}\left[1 + f(m,n) + 2\sqrt{f(m,n)}\cos\varphi\right], \qquad (4)$$

where $\rho = \eta/\hbar\omega$, $\eta$ is the quantum efficiency of the CCD camera, $\hbar\omega$ is the photon energy and $\tau$ is the exposure time. $\varphi$ represents the phase angle between the illumination and scattered optical fields. In conventional case of equal optical paths for these optical fields, $\varphi = \pi/2$, in virtue of an additional phase shift of $\pi/2$ of the non-scattered optical field [6]. Therefore, the 3$^{\text{rd}}$ term in the square brackets vanishes. We impose two constrains on the object discrete function $f$: $0 \le f \le 1$ implying that it is a passive attenuator plate, and $\sum_{m,n} f(m,n) = MN\alpha$, where $\alpha$ is a factor that accounts for diffraction efficiency of the object and collection efficiency of the optical system ($\alpha \ll 1$). For the reference purpose useful for extension of our analysis to scattering objects, $f$ can be re-expressed in terms of a scattering discrete function $s$, $f(m',n') = s(m',n')/(\delta x' \delta y')$, $\delta x' \delta y'$ being the single pixel area. The scattering function normalization is defined as: $\sum_{m,n} s(m,n) = MN \int_{\Omega_{col}} (d\sigma_{sc}/d\Omega) d\Omega$, where $d\sigma_{sc}/d\Omega$ is the differential scattering cross-section, $\Omega$ is the solid angle, and $\Omega_{col}$ is the solid angle subtended by the entrance pupil of the imaging system at the object plane origin.

Under typical conditions of imaging biological matter, the signal contains weak scattering contributions from small objects, e.g. cell nuclei, which are situated on an intense background produced by large-scale structures, e.g. cell layout, collagen bundles, etc. The visibility of small objects diminished due to the obscuring effects of large structures. In order to model this typical imaging layout, we introduce a composite object discrete function, as a sum of two Gaussian discrete functions of broad and narrow widths:

$$f(m',n') = \alpha_b \exp\left[-\frac{(m'-M/2)^2 + (n'-N/2)^2}{2\Delta_b^2}\right] + \alpha_n \exp\left[-\frac{(m'-M/2)^2 + (n'-N/2)^2}{2\Delta_n^2}\right], \tag{5}$$

where $\alpha_b, \alpha_n$ denote weighting coefficients, and $\Delta_b, \Delta_n$ denote the half width at $1/e$ fall-off of the broad and narrow discrete object functions, respectively. The object array indexing commences at the bottom right corner of the field of view. The desirable signal term (narrow Gaussian peak) associated with a small object, is situated on the background pedestal of the broad Gaussian peak, which is associated with large-scale structures. Note, the maximum amplitude of the signal $\phi_{BF}$ is limited by the saturation power $P^{sat}$, at a single pixel determined by the well depth of the CCD sensor: $\phi^{sat} = \rho \tau P^{sat}$.

Now, we turn to consider the DFM optical configuration, as shown in figure 2(b). Here the illumination power $P$ is assumed in the object arm. The object is situated at the front focal plane of the lens $L_F$, which contains a CCD sensor at its back focal plane. In the absence of the reference optical field, the optical field intensity distribution at the CCD sensor represents the normalized Fourier transform $F(m_t, n_t)$, $1 \leq m_t \leq 2M$, $1 \leq n_t \leq 2N$, of the disturbance of the optical field immediately after the object, i.e. $f(m',n')$. Therefore, the object function sampled by the CCD camera, as in the dark-field microscopy, relates to the reference-free DFM signal $F(m_t, n_t)$ via a 2D-discrete Fourier transform (DFT): $f(m,n) = DFT\{wF(m_t, n_t)\}$, where $w$ is a factor ensuring that the optical intensity integrated over the CCD areas is equal for both considered configurations:

$$\sum_{m_t, n_t} F(m_t, n_t) = \sum_{m,n} f(m,n). \tag{6}$$

In order to aid the comparison of the bright-field microscopy with DFM, the CCD sensor size has been chosen as $2M \times 2N$ to match the size of a reconstructed array of the DFM signal to that of the sampled object function.

As an example, if $f(m, n) = \exp\left[-\dfrac{(m - M/2)^2 + (n - N/2)^2}{2\Delta^2}\right]$, then,

$$wF(m_t, n_t) = \frac{16\pi^2\Delta^4}{MN} \exp\left[-2\pi^2\Delta^2\left(\frac{(m_t - M)^2}{M^2} + \frac{(n_t - N)^2}{N^2}\right)\right].$$

In DFM, the reference wave represents a plane wave of power $P$ that fills the entire sensor uniformly. The reference wave origin is assumed to be a point-like source. The interference between the reference and object optical fields is sampled by the CCD camera yielding the following results for the composite object:

$$\phi_{DFM}(m, n) =$$

$$\rho\tau \frac{P}{MN}\left\{1 + \frac{4\pi^2}{MN}\left(\alpha_n\Delta_n^4 F_n + \alpha_b\Delta_b^4 F_b\right)\right.$$

$$\left. + \frac{4\pi}{\sqrt{MN}}\sqrt{\alpha_n\Delta_n^4 F_n + \alpha_b\Delta_b^4 F_b}\,\cos\left[2\pi\left(\frac{m_0'm}{M} + \frac{n_0'n}{N}\right)\right]\right\} \qquad , \qquad (7)$$

where $(m_0', n_0')$ are indices of the discrete object function centre given as $(M/2, N/2)$. In this study, reference point-like source indices are given as $(1, 1)$. The third term in the brackets of right-hand side of Eq. (7) describes the signal term of the DFM system. Performing DFT on this signal term will fully reconstruct the object function and its lateral position on the object plane.

**b. Numerical Simulations**

We have simulated the composite object function numerically by substituting the following parameters into Eq. (5): $M = N = 1024$, $\alpha_b = \alpha_n = 1$, $\Delta_b = 700$, $\Delta_n = 10$. The discrete object function profile taken along its radial line is plotted in figure 3(a), which also serves as a representation of the recorded bright-field image under assumption of the negligible contribution of the illumination background. A 2D-bright-field image is also shown in figure 4(a). Power spectrum amplitude of the bright-field profile is presented in figure 3(b). This spectrum exhibits a large central peak associated with the broad object component, and weak sidelobes. The contrast of the desirable narrow-width signal is estimated as a ratio of the total signal maximum to the mean signal background, i.e. $C_{BF} = 90/72 = 1.25$.

In the DFM case [see figure 3(d)], the calculated signal profile is similar to the power spectrum amplitude profile of the bright-field signal [see figure 3(b)], but bears an important difference: the central peak amplitude is considerably lower. To understand it, recall that the small-angle optical scatter is tightly localized at the transfer plane contributing to the central peak amplitude, as shown in figure 3(b). In the DFM case, though, the central peak is clipped, or effectively attenuated with respect to the high spatial frequency components due to the CCD's saturation limit, whereas the wide-angle optical scatter is unaffected. When the DFM radial signal profile is reconstructed by performing a DFT [figure 3(c)], the power spectrum amplitude of the broad object function, that represent the background, is greatly diminished, whereas the narrow object function remains unaltered. The net result demonstrates the contrast enhancement of the small objects $C_{DFM} = 25/3 = 8.3$, which gives a contrast improvement of 6.7 in comparison with that of the bright-field microscopy.

It is also worthwhile to compare image contrast achievable using DFM with another well-known technique, dark-field microscopy. In the dark-field microscopy, an intense central peak at the transfer plane due to the small-angle optical scatter is completely removed by placing a field stop [see figure 3(f)], on axis, in front of the CCD camera. The profile of the simulated signal after removal of the central component at the transfer plane is shown in figure 3(e). However, in contrast with the DFM, the dark-field microscopy signal exhibits side-lobe artefacts due to the rectangular function filter response of the beam stop.

In order to aid visual perception of our results, we have computed and plotted pseudo colour images of the composite object obtained using: bright-field, DFM, and dark-field microscopy. A bright-field image with a broad background component is shown in figure 4(a). A dramatic increase in image contrast can be seen in the DFM and dark-field microscopy images shown in figures 4(b) and (c), respectively. Although some artefacts are introduced in DFM, they are not as apparent as those in the dark-field microscopy. Furthermore, sidelobes in DFM can be removed by apodization, which is considerably more difficult to implement in hardware, as in the case in the dark-field microscopy.

## 3. Experiment

We have carried out experimental investigation of the imaging performance of the DFM system in transmission and reflection-modes. Two schematic diagrams of the DFM transmission and reflection experimental systems are shown in figures 5(a) and (b), respectively. Transmission-mode is commonly used for *in vitro* microscopic observation of thin biological samples, e.g. cell cultures, while reflection-mode is important for imaging of relatively translucent thick samples, or, presumably, *in vivo* imaging of low-scattering biological samples.

In transmission-mode DFM, an optical output from a helium-neon laser ($\lambda = 632.8$ nm, $P = 5$ mW) was equally split into object and reference arms. In the object arm, a sample was placed at the primary objective plane in the path of the collimated laser beam, where light was either scattered or transmitted through the sample. An assembly of an infinity-corrected objective lens $L_{obj}$ (Olympus, $\times 40$, NA = 0.65), and a relay lens $L_r$ (achromat-doublet, focal length $f_r = 60$ mm), produced a magnified image of the object at the secondary object plane (see figure 5). An aperture was placed at this plane to limit the field of view of the overall imaging system, and in order to avoid aliasing in the reconstructed image. The magnified image was projected onto the transfer plane by a Fourier lens $L_F$ (achromat-doublet, focal length $f_F = 100$ mm), where a CCD camera (Photometrics, CoolSNAP *cf*, 12-bit, 1392×1040) was situated. In the reference arm, the laser light was coupled into a single-mode optical fibre (mode-field diameter 4.0 µm at 632.8 nm, NA = 0.11). The distal end of the fibre was placed off-axis at the object plane. The interference between the optical field due to the magnified image of the object and the quasi plane reference field combined at the transfer plane was recorded by the CCD camera.

In order to test the basic performance of DFM, as applied to imaging of biological objects, we have fabricated a biological phantom. To mimic cell nuclei suspension in the cytoplasm matrix, polystyrene spheres were embedded into the polyvinyl alcohol matrix (PVA). The spheres were provided as aqueous suspensions (ProSciTech, 5.0 % w/v in water). 0.5 and 2.0 mL of the 2- and 5- µm -diameter sphere suspensions were combined with 10 % w/v PVA solutions in deionised water. Thin solid films where made by spin coating this mixture onto quartz cover slips at 700 revolutions per minute for 10 seconds [1]. As a result, a sample of well-separated non-aggregated sphere mixture was produced. The polystyrene spheres relative refractive index was estimated to be equal to 1.05, which was very close to that of the typical cellular nuclei, i.e. 1.06. A Fourier hologram of this sample was recorded by the CCD camera using exposure time of 1 ms for each image. For alignment and initial observations, RS Image software by Roper Scientific was used to operate the camera, and display captured images. The digital data transfer from the CCD to a personal computer was computer-controlled from the LabVIEW platform. Image processing, storage, and display were implemented in Matlab.

The diagram of the DFM experimental setup in reflection-mode is shown in figure 5(b). This configuration is similar to that of the transmission-mode, except the direction of the illumination beam incidence on the sample is reversed. In our case, the illumination beam was, first, expanded by a telescope [not shown in figure 5(b)] and then introduced into the system via a pellicle beamsplitter, which was located between the object plane and $L_r$. This configuration ensured that a beam incident on the sample rendered collimated after passing the objective lens. In reflection-mode, light backscattered from the object is collected via the same optical train as in transmission-mode.

In order to carry out comparative study of imaging contrast of DFM versus bright-field microscopy, DFM was operated in reflection-mode. The bright-field microscopy system was reconfigured from the DFM system by increasing the separation from the object plane to $L_F$, and from $L_F$ to the CCD camera, from $f_F$ to $2f_F$, such that an image was recorded directly onto the CCD, in similarity with the principal configuration shown in figure 2(b). Equal illumination was ensured in both configurations. The illumination beam reflected from a sample interface served as a background optical field of the bright-field microscopy system.

## 4. Results

Figure 7 present results of experimental observations of image contrast enhancement of the DFM technique. Images obtained using DFM and bright-field microscopy systems are presented in figure 6(a) and (b), respectively. In bright-field microscopy, gross morphology of the surrounding polyvinyl alcohol, is predominantly visible. As a result, small objects, e.g. 2-$\mu$m polystyrene spheres, are largely obscured. This effect is considerably improved in case of DFM. For example, a 2-$\mu$m sphere, marked by an arrow A, not clearly observable on the bright background in the bright-field image [figure 6(b)], is visible in the DFM image [figure 6(a)]. Also, three 2-$\mu$m spheres, marked by an arrow B, appear to be well resolved in DFM [figure 6(a)], while their separate appearance is somewhat ambiguous in the bright-field microscopy. Comparison of these two images shows the efficiency of DFM for image contrast enhancement, as applied to small-size objects.

Next, we demonstrated the use of optical scatter for the DFM filtering. A DFM image captured in transmission-mode is shown in figure 7(a). This image features a noticeable background on which 2- and 5-$\mu$m polystyrene spheres are observed. Slightly defocused appearance of these spheres (surrounded by rings) is attributed to the deliberate defocusing of the imaging optics in order to reduce the blooming effect of the tightly focussed illumination beam on the CCD camera.

In order to demonstrate the selective enhancement of a small object of interest, a suitable digital mask was required. This pick up mask was designed by processing the recorded optical scatter signal of the selected object. In our case, the optical scatter represented an envelope of the Fourier hologram of this object. In order to obtain this envelope, an individual scatterer of e.g. 2-$\mu$m in diameter was cropped out from the reconstructed image, for example a sphere marked "2-$\mu$m" in figure 7(a). The discrete crop-out function was chosen to be a 2D-Gaussian-shape array that was centred at the selected sphere and padded with zeros elsewhere to eliminate contributions from the other objects. The product of the reconstructed image and crop-out function produced an image of the selected sphere and its vicinity on a dark background. A DFT of the resultant array followed by the envelope detection yielded the optical scatter of the selected sphere. This mask design process was repeated for two different 2-$\mu$m-spheres, and also 5-$\mu$m spheres. Two envelopes of each size were averaged to produce masks for 2- and 5-$\mu$m spheres, which are presented in figures 8(a) and (b), respectively. As expected, the optical scatter pattern of 5-$\mu$m-sphere exhibits a small central lobe surrounded by, at least, one ring, whereas the optical scatter of 2-$\mu$m sphere comprises one large-size central lobe. The comparison of these optical scatters with optical scatters calculated using the Mie scattering theory has been reported to show reasonable agreement [1]. These optical scatter patterns represent the basis for digital mask design for the purpose of Fourier filtering of an object of interest.

An efficient method to discriminate an object of one kind (e.g. 5-$\mu$m polystyrene sphere) in favour of an object of the other (e.g. 2-$\mu$m polystyrene sphere) is to multiply the ratio of the two normalised optical scatters with the original hologram. The ratio of the 2-$\mu$m to 5-$\mu$m normalised optical scatters, a pick up mask, is shown in figure 8(c). This mask takes the form of a doughnut and features a high attenuation zone at the centre with a high transmission wing that fades off outwards of the mask. This design picks up wide-angle scatter contributions, which are pronounced in the case of 2-$\mu$m spheres. Most of the signal in the hologram that contributes to the reconstruction of 5-$\mu$m spheres is localized in the high attenuation zone of the pick-up mask and, therefore, this signal is suppressed, rendering the 5-$\mu$m spheres with dim appearance. When the product of the pick-up mask and original DFM hologram is reconstructed, the predominant enhancement of the 2-$\mu$m spheres is clearly observable, as shown in figure 7(b). The image background is also largely suppressed.

## 5. Discussion

As we have demonstrated in the preceding sections, DFM is not only capable of producing images via DFT reconstruction, but is also capable of producing optical scatter imaging, providing extra information of sample properties. For example, a small aperture and non-absorbing sphere of the same diameter would produce an almost identical image, although, the optical scatter would be different for these two objects. Digital Fourier microscopy is a suitable technique to pick up this difference. Similarly, two equal-size spheres of different refractive indices would be distinguished using DFM, even though their images should be identical. It is important to point out that since DFM data is stored in digital format, the optical scatter can be retrieved at the post-processing stage. This renders the technique as flexible, as to that of the previously reported (OSI) [4] in terms of the optical scatter control for the enhancement of features of interest. Exploiting the flexibility of digital filtering, differentiation between various objects of different optical form-factors can be made possible using DFM, one example being spherical and spheroidal objects.

It is worthwhile to comment on several problems of DFM. Since the core recording configuration of DFM is Fourier holography, it suffers from a drawback common to Fourier holography: a collimated illumination light beam in the object arm (as well as light scattered from large structures) is tightly focussed into an intense spot on the CCD camera. This causes saturation of its several central pixels. As a consequence, the CCD gain had to be adjusted to prevent photo-damage and blooming effects. The sensitivity to detection of low-intensity high-spatial frequency scatter due to small objects appears to be compromised. In order to ameliorate these undesirable effects, a CCD camera can be slightly displaced from the transfer plane, so that the illumination light is no longer tightly focused. The sensitivity of DFM to small objects can, therefore, be considerably improved, at the expense of slight degradation of the lateral resolution of the system [15]. In reflection-mode, the tight focussing effect is exacerbated by the presence of specular reflections from the multiple optical elements in the objective lens and, also, from the air/sample interface. Even though a specular reflection from each interface represents only several per cent of the total incident optical power, the backscatter efficiency is also reduced (several orders of magnitude for micron-size sphere of relative refractive index of 1.06, as a nucleus in cytoplasm), as compared to that of the forward scatter. Therefore, prevention of the tight focussing on the CCD camera presents a problem in DFM, and warrants further study.

There is an ample scope for further improvement of the DFM performance. First of all, the optical train of elements following the objective lens was not optimized, and it accounted for the suboptimal image quality presented in this paper. Secondly, a number of optical elements of conventional surface quality and cleanliness inevitably resulted to the excess noise in the recorded holograms. The incoherent contribution to the noisy background of the recorded hologram can be compensated by employing the phase shifting methods [17]. Imaging of highly scattering samples is expected to present a problem for DFM, since a high-coherence illumination source is employed in this technique. A parasitic interference originated from the multiple-scattering within these samples will degrade the reconstructed image quality significantly.

To illustrate the significance of the reported DFM technique, consider recording of the Fourier hologram of the cell culture conventionally modelled as a set of nuclei (spheroids) embedded in the lower refractive index cytoplasm. It is expected that adaptive digital filtering followed by digital image reconstruction enhances spheroids that match the pre-selected parameters of the digital mask, with the most important parameter being spheroid size. The application of a set of digital masks computed for a range of spheroid sizes can potentially yield the size distribution of the cell nuclei, which will serve as the key diagnostic discriminate of, for example, cancer [18].

## 6. Conclusion

In conclusion, we have reported on advances in digital Fourier microscopy, a holographic implementation of the optical scatter imaging. This DFM approach is based on the application of digital Fourier holography to recording of comprehensive information of the object, including optical scatter of the object constituents. This approach confers several merits, as compared to optical scatter imaging implemented in hardware: objects of interest have been demonstrated to be enhanced by applying digital filtering to Fourier holograms recorded using DFM. Contrast of small-size features on the slowly-varying background has been demonstrated theoretically and experimentally to be enhanced, in similarity with the dark-field microscopy. We believe that DFM is particularly suitable for *in vitro* applications for imaging of thin biological objects, as it provides an efficient discrimination mechanism for biological constituents of interest.

**Figures and Captions**

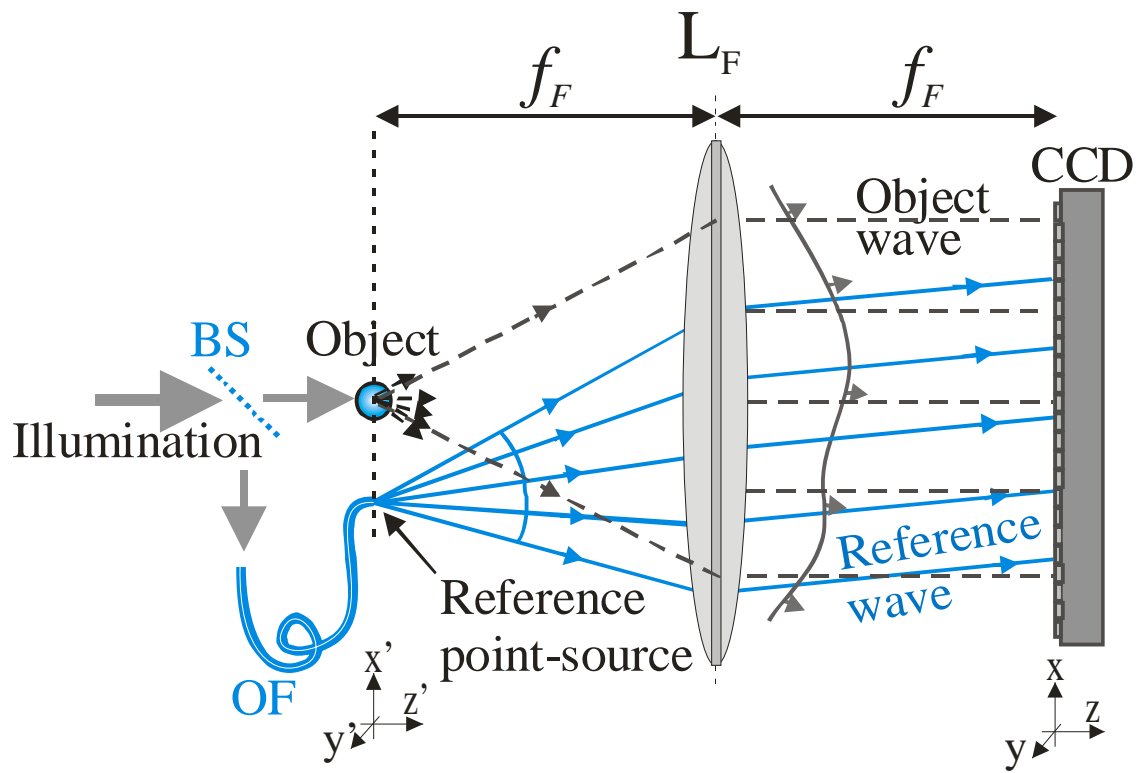

**Figure 1.** Schematic diagram of a DFM. A Fourier lens L$_F$, transforms the optical field at the object plane onto the transfer plane, where it interferes with the reference field and is recorded by a CCD. An optical fibre OF, distal end represents the reference point-source also located at the object plane. BS, beam splitter.

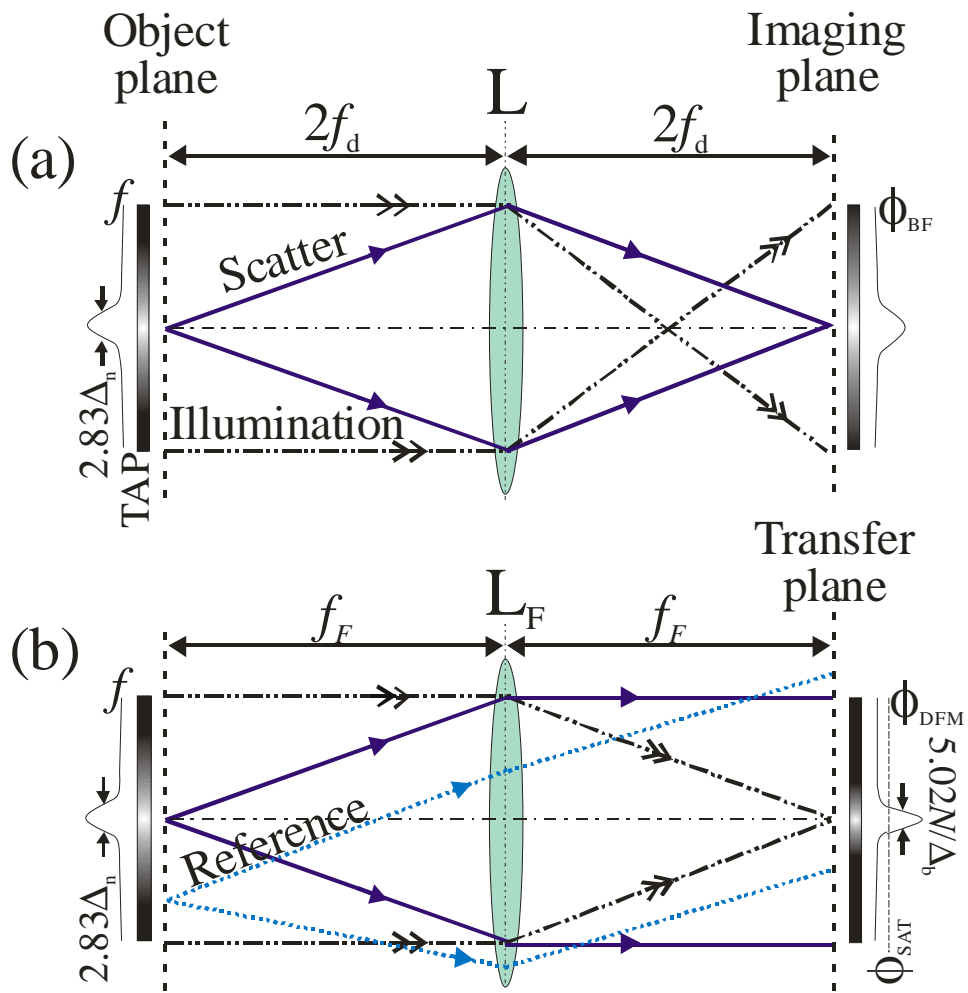

**Figure 2.** Ray diagrams of (a) bright-field and (b) DFM configurations. TAP, thin attenuation plate. Lenses L, L$_F$ have focal lengths of $f_d$, $f_F$, respectively. No interference fringes are shown in pictorial representation of $\phi_{DFM}$.

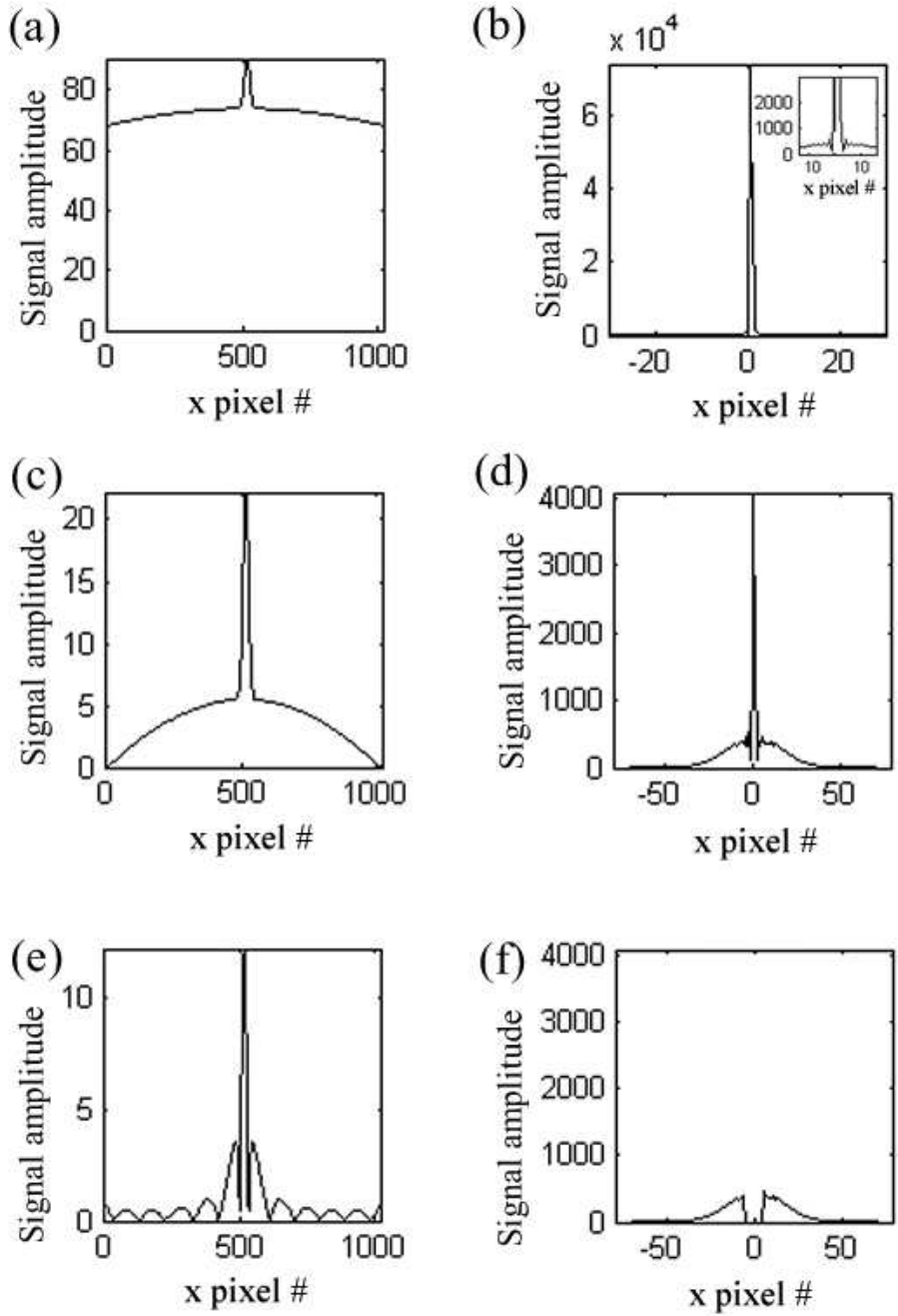

**Figure 3.** Plots of the calculated signal profiles of the composite object images along the radial line: (a) bright-field, (e) dark-field microscopy, and (c) power spectrum of the DFM signal, i.e. reconstructed signal profile. Central parts of plots of the signal profiles calculated on the transfer plans of (b) bright-, (f) dark-field microscopy, (d) DFM. An inset (b) shows details of the signal sidelobes.

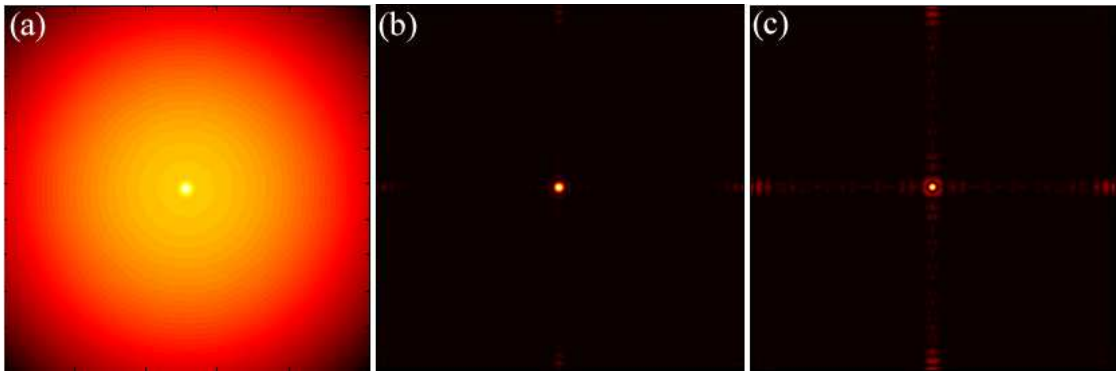

**Figure 4.** Pseudo colour images of (a) bright-field, (b) DFM, and (c) dark-field microscopy simulated numerically.

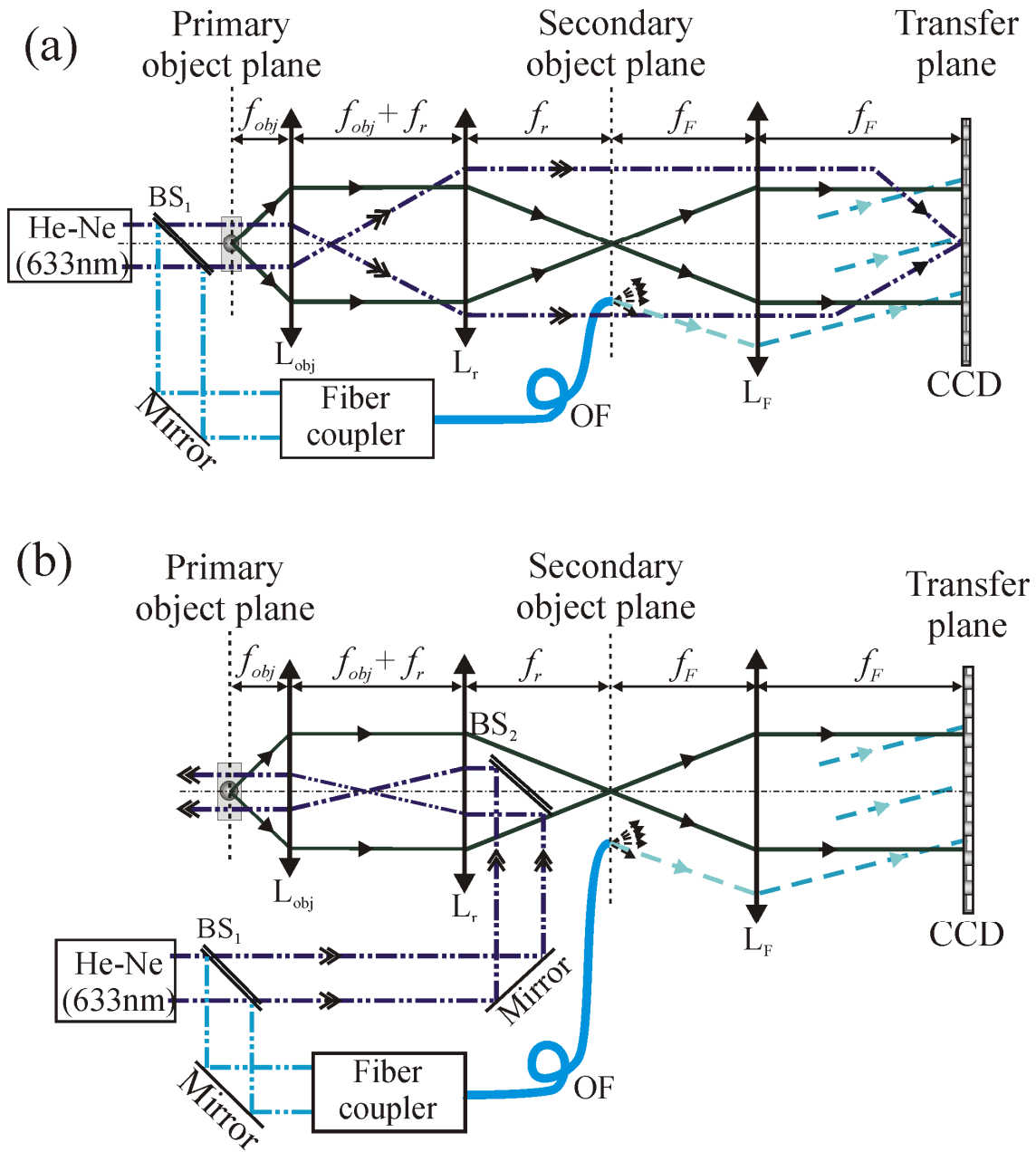

**Figure 5.** Diagrams of DFM experimental setups in (a) transmission, and (b) reflection modes. BS₁ and BS₂, non-polarising and pellicle beam splitters, respectively; $L_{obj}$, $L_r$, $L_F$ objective, relay, Fourier lenses, respectively of the corresponding focal lengths of $f_{obj}$, $f_r$, $f_F$; OF, single-mode optical fibre.

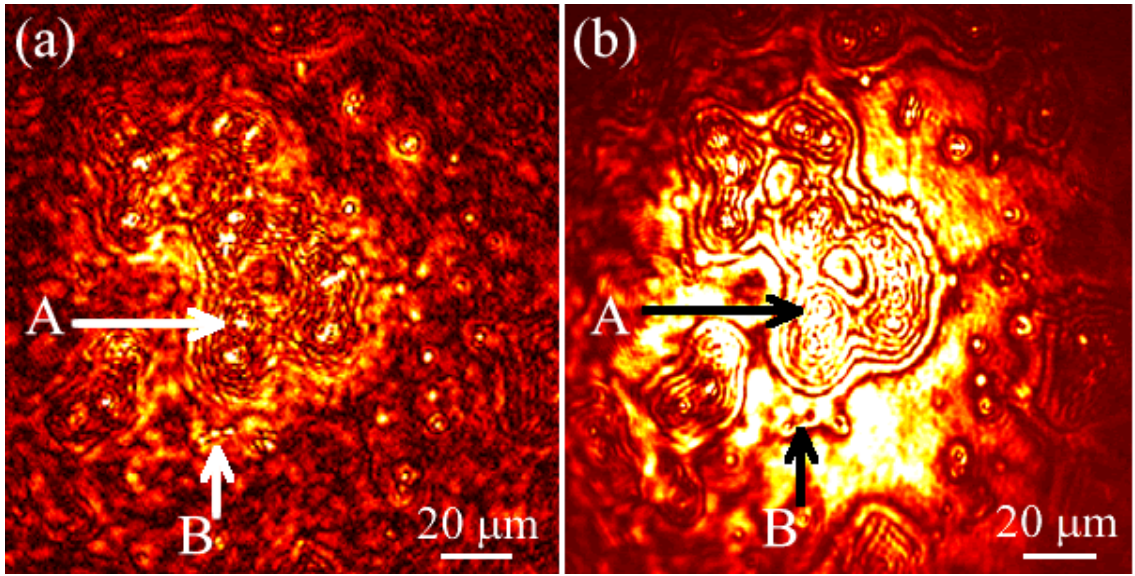

**Figure 6.** Images of the biological phantom obtained in reflection mode using (a) DFM, and (b) bright-field configurations. Arrows A and B show one and three spheroids, respectively in both image panels.

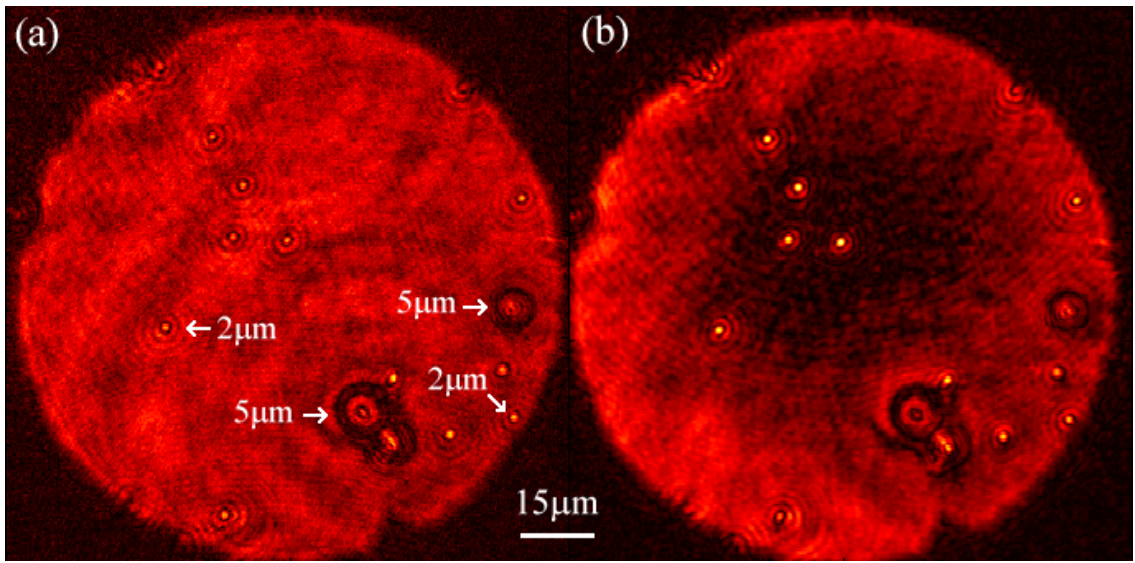

**Figure 7.** Images of the biological phantom recorded using DFM in transmission mode: (a) non-processed; (b) filtered. Note enhanced appearance of 2-µm spheres. Arrows mark 2-µm and 5-µm spheres used for the digital mask design. See text for details.

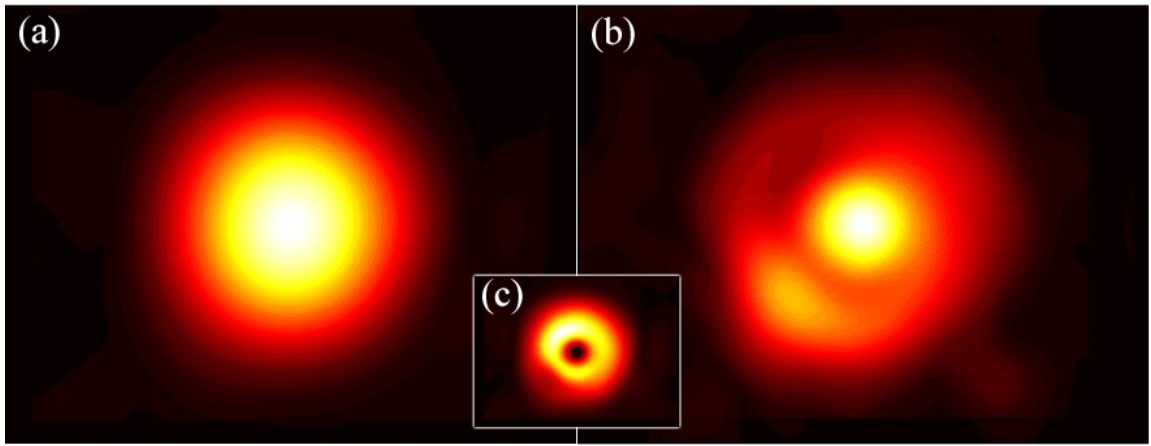

**Figure 8.** Optical scatter patterns from (a) 2- and (b) 5- μm spheres. (c) A digital mask (de-magnified), formed by the ratio of the 2- and 5- μm optical scatter patterns.